\documentclass[12pt]{iopart}

\usepackage{graphicx}
\usepackage{bm}
\def\mbf(#1){\mbox{\boldmath $#1$}}

\newcommand{\avg}[1]{\left\langle {#1} \right\rangle}

\begin{document}

\title[Electronic transport properties of graphene nanoribbons]{
Electronic transport properties of graphene nanoribbons
}

\author{Katsunori Wakabayashi$^{1,2}$, Yositake Takane$^3$, Masayuki Yamamoto$^{1}$ and
Manfred Sigrist$^4$}

\address{
$^1$International Center for Materials Nanoarchitectonics(MANA),
National Institute for Materials Science (NIMS), Namiki 1-1,
Tsukuba 305-0044, Japan,\\
$^2$PRESTO, Japan Science and Technology Agency(JST), Kawaguchi
332-0012, Japan, \\
$^3$Department of Quantum Matter, AdSM, Hiroshima University,
Higashi-Hiroshima 739-8530, Japan, \\
$^4$Theoretische Physik, ETH-Z\"urich, Z\"urich CH-8093, Switzerland.
}
\begin{abstract}
We will present brief overview on the electronic and transport properties
of graphene nanoribbons focusing on the effect of edge shapes and 
impurity scattering. 
The low-energy electronic states of graphene have two non-equivalent 
massless Dirac spectrum. The relative distance between these two Dirac
 points in the momentum space and edge states due to the existence 
of the zigzag type graphene edges are decisive to the
electronic and transport properties of graphene nanoribbons. 
In graphene nanoribbons with zigzag edges (zigzag nanoribbons),
two valleys related to each Dirac spectrum are well separated in momentum space.
The propagating modes in each
 valley contain a single chiral mode originating from a partially flat 
band at band center. This feature gives rise to a perfectly conducting
 channel in the disordered 
system, if the impurity scattering does not connect the two valleys, i.e. for
long-range impurity potentials. Ribbons with short-range impurity potentials, 
however, through inter-valley scattering 
display ordinary localization behavior. 
On the other hand, the low-energy spectrum of graphene nanoribbons with
armchair edges (armchair nanoribbons) is described as 
the superposition of two non-equivalent Dirac points of graphene. 
In spite of the lack of well-separated two 
valley structures, the single-channel transport subjected to
long-ranged impurities is nearly perfectly conducting,
where the backward scattering matrix elements in the lowest order vanish 
as a manifestation of internal phase structures of the wavefunction.
For multi-channel energy regime, however, the conventional exponential
decay of the averaged conductance occurs.
Symmetry considerations lead to the classification of disordered zigzag
 ribbons into the unitary class for long-range impurities, and the
 orthogonal class for short-range impurities. 
Since the inter-valley scattering is not completely absent, 
armchair nanoribbons can be classified into orthogonal universality
class irrespective of the range of impurities.
\end{abstract}

\maketitle

\section{Introduction}
Recently graphene, a single-layer hexagonal lattice
of carbon atoms, has emerged as a fascinating system for
fundamental studies in condensed matter physics, as well
as the promising candidate material for future application
in nanoelectronics and molecular devices.\cite{novoselov}
The honeycomb crystal structure of single layer graphene
consists of two nonequivalent sublattices and results in a unique 
band structure for the itinerant $\pi$-electrons near
the Fermi energy which behave as massless Dirac fermion.
The valence and conduction bands touch conically at two nonequivalent
Dirac points, called $\bm{K_+}$ and $\bm{K_-}$ point,
which form a time-reversed pair, {\it i.e.} opposite chirality.
The chirality and a Berry phase of $\pi$ at the two Dirac points 
provide an environment for highly unconventional and fascinating
two-dimensional electronic 
properties,\cite{neto}  
such as the half-integer quantum Hall effect,\cite{qhe} 
the absence of backward scattering,\cite{ando,ando.nakanishi}
$\pi$-phase shift of the Shubnikov-de Haas oscillations.\cite{kopelvich} 

The successive miniaturization of the graphene electronic devices inevitably
demands the clarification of edge effects on the electronic structures
and electronic transport properties of nanometer-sized graphene. 
The presence of edges in graphene has strong implications
for the low-energy spectrum 
of the $\pi$-electrons.\cite{peculiar,nakada,prb.1999} 
There are two basic shapes of edges, {\it armchair} and {\it zigzag} which
determine the properties of graphene ribbons. 
It was shown that ribbons with zigzag edges (zigzag ribbon) possess 
localized edge states with energies close to
the Fermi level.\cite{peculiar,nakada,prb.1999,Phd}
These edge states correspond to the non-bonding configurations
 as can be seen by examining the analytic solution for
semi-infinite graphite with a zigzag edge for which the wave functions of the edge states reside on
one sublattice only.\cite{peculiar}
In contrast, edge states are completely absent for ribbons with armchair edges. 
Recent experiments support the evidence of edge localized states.\cite{enoki,fukuyama}
Also, graphene nanoribbons can experimentally be produced by using
lithography techniques and chemical techniques.\cite{Kim,dai,muellen,unzipcntube,mitribbon}

The electronic transport through graphene nanoribbons shows 
a number of intriguing phenomena 
such as zero-conductance Fano resonances,\cite{prl,ijmpb}
valley filtering,\cite{rycerz}, half-metallic conduction\cite{son},
spin Hall effect,\cite{kane}, and
perfectly conducting channel.\cite{prl2007}
Recent studies also clarify the unconventional transport through
graphene junctions, quantum point contact and heterojunctions.\cite{ijmpb,sns,rmp.bee,akhmerov,qpc,klein,nakabayashi,cresti,nikolic,abanin,morooka,peres2009,ryzhii,williams,molitor,dai.transistor,tapaszto,stampfer,chen,campos,miyazaki}
It is also expected that 
the edge states play an important role for the magnetic properties
in nanometer-sized graphite systems,
because of their relatively large contribution to the density of
states at the Fermi
energy.\cite{peculiar,prb.1999,rpa,kusakabe,moebius,yamashiro,harigaya,yoshioka,hikihara,palacios,fernandes-rossier,sasaki,kumazaki,hod} 
Recent studies explore the robustness of edge states to the size and
geometries\cite{hod,ezawa,kundin}, and various edge structures and modification.\cite{hod,mauri}

Since the graphene nanoribbons can be viewed as a new class of 
quantum wires, one might expect that random impurities 
inevitably cause Anderson localization, i.e.
conductance decays exponentially with increasing system length $L$
and eventually vanishes in the limit of $L \to \infty$.
However, it was shown that carbon nanotubes with long-ranged impurities
possess a perfectly conducting channel.\cite{suzuura} 
Recent studies show
that perfectly conducting channels can be stabilized in two
standard universality classes.
One is the symplectic universality class with an odd number of conducting
channels,\cite{suzuura,takane1,takane2}
and the other is the unitary universality class with the imbalance between
the numbers of conducting channels in two propagating
directions.~\cite{prl2007,ohtsuki,takane4}
The symplectic class consists of systems having time-reversal symmetry
without spin-rotation invariance, while the unitary class is characterized
by the absence of time-reversal symmetry.~\cite{beenakker.rmt}

In this paper, we will give a brief overview on the electronic transport
properties of disordered graphene nanoribbons. In zigzag nanoribbons,
the edge states play an important role, since they appear as special modes with partially flat bands
and lead under certain conditions to chiral modes separately in the two valleys. There is one such mode of opposite orientation 
in each of the two valleys of propagating modes,
which are well separated in $k$-space. The key result of this study is that for disorder
without inter-valley scattering a single perfectly conducting channel emerges
introduced by the presence of these chiral modes. This effect disappears
as soon as inter-valley scattering 
is possible. 
On the other hand, the low-energy spectrum of graphene nanoribbons with
armchair edges (armchair nanoribbons) is described as 
the superposition of two non-equivalent Dirac points of graphene. 
In spite of the lack of well-separated two 
valley structures, the single-channel transport subjected to
long-ranged impurities is nearly perfectly conducting,
where the backward scattering matrix elements in the lowest order vanish 
as a manifestation of internal phase structures of the wavefunction.\cite{yamamoto}
For multi-channel energy regime, however, the conventional exponential
decay of the averaged conductance occurs.
Symmetry considerations lead to the classification of disordered zigzag
 ribbons into the unitary class for long-range impurities, and the
 orthogonal class for short-range impurities. 
Since the inter-valley scattering is not completely absent, 
armchair nanoribbons can be classified into orthogonal universality
class irrespective of the range of impurities.

\section{Electronic states of graphene and nanoribbons}
\subsection{Tight-binding model and edge states}
There are two typical shapes of a graphene edge,
called {\it armchair} and {\it zigzag}.
The two edges have 30 degrees difference in their
cutting direction.
Here we briefly discuss the way that the graphene edges
drastically change the $\pi$ electronic structures.\cite{peculiar}.
Especially, a zigzag edge provides the localized edge state,
while an armchair edge does not show such localized states.

\begin{figure}
\begin{center}
\includegraphics[width=0.9\textwidth]{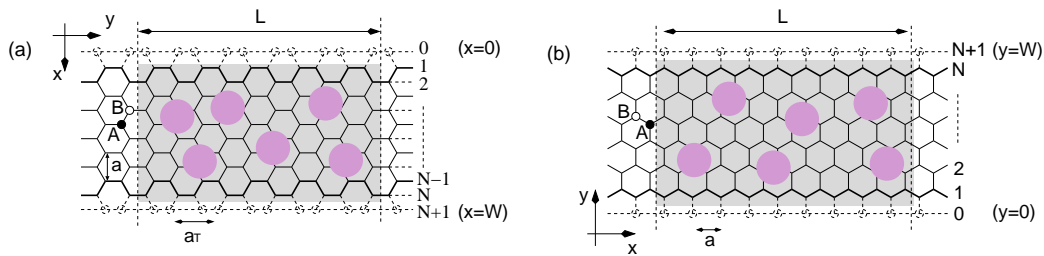}
\end{center}
\caption{Structure of graphene nanoribbon with (a) armchair edges
(armchair ribbon) and (b) zigzag edges (zigzag ribbon). 
The lattice constant is $a$ and $N$ defines the ribbon
width. The circles with dashed line indicate the missing carbon
atoms for the edge boundary condition of massless Dirac equation.
The disordered region with randomly distributed impurities lines in the
 shaded region and has the length $L$(see text in section 3).
Randomly distributed circles schematically represent the long-ranged impurities. 
}
\label{fig:graphite-edge}
\end{figure}

A simple and useful model to study the edge and size effect 
is one of the graphene ribbon models as shown in
Figs.~\ref{fig:graphite-edge}(a) and (b).
We define the width of graphene ribbons as $ N $, 
where $ N $ stands for the number of the dimer (two carbon sites) lines
for the armchair 
ribbon and by the number of the zigzag lines for the zigzag ribbon,
respectively.  
It is assumed that all dangling bonds at graphene edges are 
terminated by hydrogen atoms, and thus give no contribution to 
the electronic states near the Fermi level.
We employ a single-orbital tight binding model for the
$\pi$ electron network.
The Hamiltonian is written as,
\begin{equation}
H = - t\sum_{\langle i,j \rangle} 
      c^\dagger_ic_j
    + \sum_i V_i      c^\dagger_ic_i,
\label{eq1:hamil}
\end{equation}
where the operator $c^\dagger_i$  creates an $\pi$-electron on the site $i$.
$\langle i,j \rangle$ denotes the summation over the nearest neighbor
sites. 
$ t $ the transfer integrals between all the nearest neighbor sites
are set to be unity for simplicity. 
This is sufficient to show the intrinsic 
difference in the electronic states originating from the topological 
nature of each system. 
The value of $ t $ is considered to be about 
$2.75$eV in a graphene system. 
The second term in Eq. (\ref{eq1:hamil}) represents the 
impurity potential, 
$V_i=V(\bm{r}_i)$ is the 
impurity potential at a position $\bm{r}_i$.
The effect of impurity potential on the electronic transport properties
will be discussed in the next section. 

\begin{figure}
\begin{center}
\includegraphics[width=0.9\linewidth]{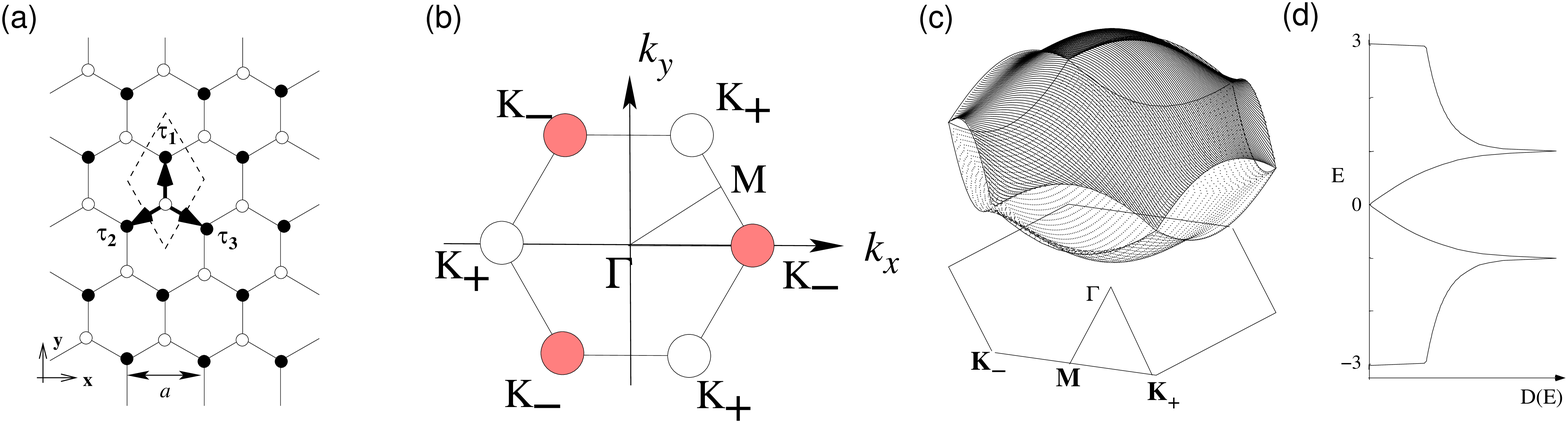}
\end{center}
\caption{ 
(a) Graphene sheet in real space, where the black (white) circles
mean the A(B)-sublattice site. $a$ is the lattice constant.
Here $\bm{\tau_1}=(0,a/\sqrt{3})$, $\bm{\tau_2}=(-a/2,-a/2\sqrt{3})$, and
$\bm{\tau_3}=(a/2,-a/2\sqrt{3})$.
(b) First Brillouin zone of graphene. 
$\bm{K_+}=\frac{2\pi}{a}(\frac{1}{3},\frac{1}{\sqrt{3}})$,
$\bm{K_-}=\frac{2\pi}{a}(\frac{2}{3},0)$, $\bm{\Gamma}=(0,0)$
(c) The $\pi$ band structure and (d) the density of
states of graphene sheet.
The valence and conduction bands make contact at the degeneracy
point $\bm{K_\pm}$.}
\label{fig:gsheet}
\end{figure}

Prior to the discussion of the $\pi$-electronic states of graphene nanoribbons,
we shall briefly review the $\pi$-band structure of a graphene
sheet\cite{wallace}.
To diagonalize the Hamiltonian for a graphene sheet, we use a basis
of two-component spinor, 
$\mbf(c)^\dagger_{\bf k}=$($c^\dagger_{A\bf k}$, $c^\dagger_{B\bf k}$),
which is the Fourier transform of 
($c^\dagger_{i\in A}$, $c^\dagger_{i\in B}$).
Let $\mbf(\tau_1)$, $\mbf(\tau_2)$, $\mbf(\tau_3)$ be the displacement vectors
from a B site to its three nearest-neighbor A sites, defined so
that $\mbf(\hat{z}\cdot \tau_1 \times \tau_2)$ is positive (Fig.2(a)). $\mbf(\hat{z})$ is
the normal vector to the graphene sheet.
In this representation, the Hamiltonian is written as $H=\sum_{\bf
k}{\bf c}^\dagger_{\bf k}H_{\bf k}{\bf c}_{\bf k}$ and
\begin{equation}
H_{\bf k} = -t\sum_{i=1}^3\left(
\cos(\mbf(k\cdot \tau_i)) \hat{\sigma}_x +
\sin(\mbf(k\cdot \tau_i)) \hat{\sigma}_y
\right),
\end{equation}
where $\hat{\mbf(\sigma)}=(\hat{\sigma}_x, \hat{\sigma}_y,
\hat{\sigma}_z)$ are the Pauli matrices. 
Then, the energy eigenvalues are 
$E^\pm_{\bf k} = \pm t|\sum_{i=1}^3 \exp(\mbf(k\cdot \tau_i))|$.
Since one carbon site has one $\pi$-electron on average,
only $E_{\bf k}^{-}$-band is completely occupied.

In Figs.\ref{fig:gsheet} (b)-(d), 
the 1st Brillouin Zone (BZ) of graphene lattice, 
the energy dispersion of $\pi$-bands in the 1st Brillouin Zone (BZ) 
and the corresponding density of states are depicted, respectively.
Near the $\Gamma$ point, 
both valence and conduction bands have 
the quadratic form of $k_x$ and $k_y$, {\it i.e.} $E_{\bm k}=\pm
(3-3|{\bm k}|^2/4)$.
At the M points, the middle points of sides of the hexagonal
BZ, the saddle point of energy dispersion appears 
and the density of states diverges
logarithmically.
Near the K point of the corner of hexagonal 1st BZ,
the energy dispersion is linear in 
the magnitude of the wave vector,
$E_{\bf k} = \pm  \sqrt{3}ta |\mbf(k)|/2$,
where the density of states linearly depends on energy.
Here $a(=\sqrt{3}|\mbf(\tau_i)|(i=1,2,3))$ is the lattice constant. 
The Fermi energy is located at the K points and
there is no energy gap at these points,
since  $E_{\bf k}$  vanishes at these points
by the hexagonal symmetry.

\begin{figure}
\begin{center}
\scalebox{.28}{\includegraphics{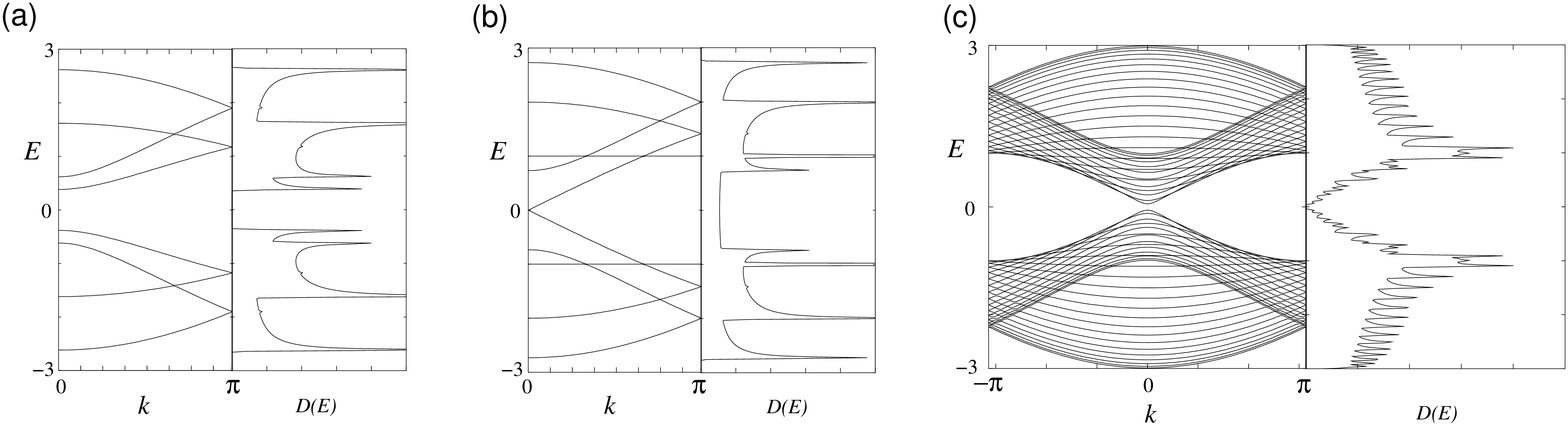}}
\end{center}
\caption{
Energy band structure $E(k)$ 
and density of states $D(E)$ of armchair ribbons of
various widths [(a) $N=4$, (b) $5$ and (c) $30$ ].}
\label{fig:armchair}
\end{figure}

The energy band structures of armchair ribbons
are shown in Figs.\ref{fig:armchair} (a)-(c), for three different ribbon
widths, together with the density of states.
The wave number $k$ is normalized by the length of the primitive
translation vector of each graphene nanoribbon, and the
energy $E$ is scaled by the
transfer integral $t$.
The top of the valence band and the bottom of the conduction band are
located at $k=0$.
It should be noted that
the ribbon width decides whether the system is
metallic or semiconducting . As shown
in Fig.\ref{fig:armchair} (b), the system is metallic when $N=3M-1$,
where $M$ is an integer.
For the semiconducting ribbons, the direct gap decreases with increasing
ribbon width and tends to zero in the limit of very large $N$.
For narrow non-doped metallic armchair nanoribbons, the energy gap can acquire due to Peierls
instabilities toward low-temperatures,\cite{igami} which is consistent with
the recent density functional theory calculation.\cite{hod,son2}

\begin{figure}
\begin{center}
\scalebox{.28}{\includegraphics{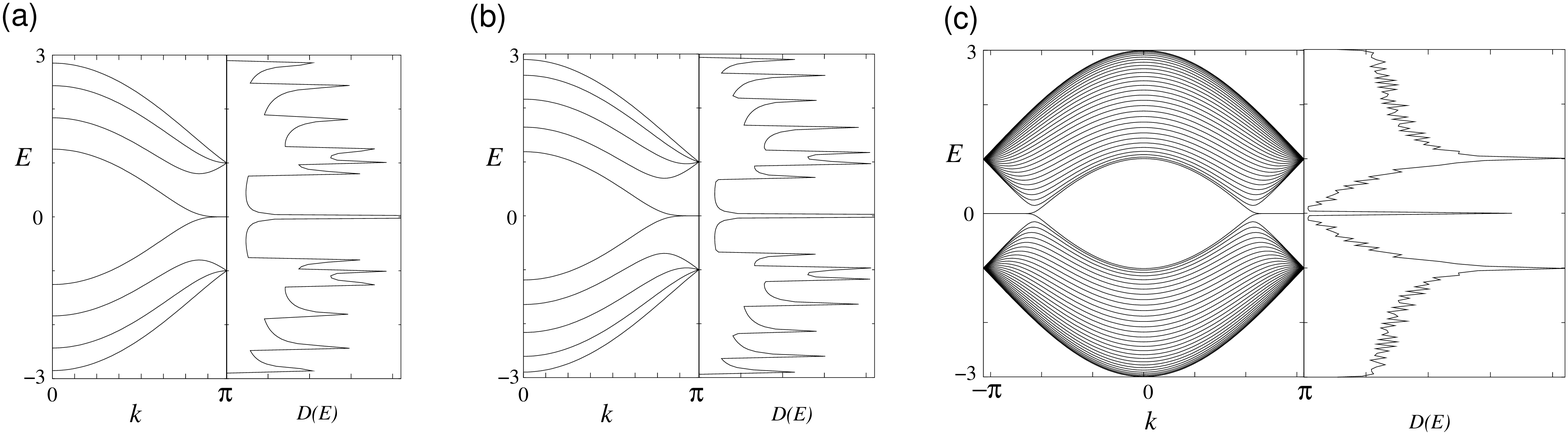}}
\end{center}
\caption{
Energy band structure $E(k)$ 
and density of states $D(E)$ 
of zigzag ribbons of
various widths [(a) $N=4$, (b) $5$ and (c) $30$ ].}
\label{fig:zigzag}
\end{figure}

\begin{figure}[h]
\begin{center}
\scalebox{.45}{\includegraphics{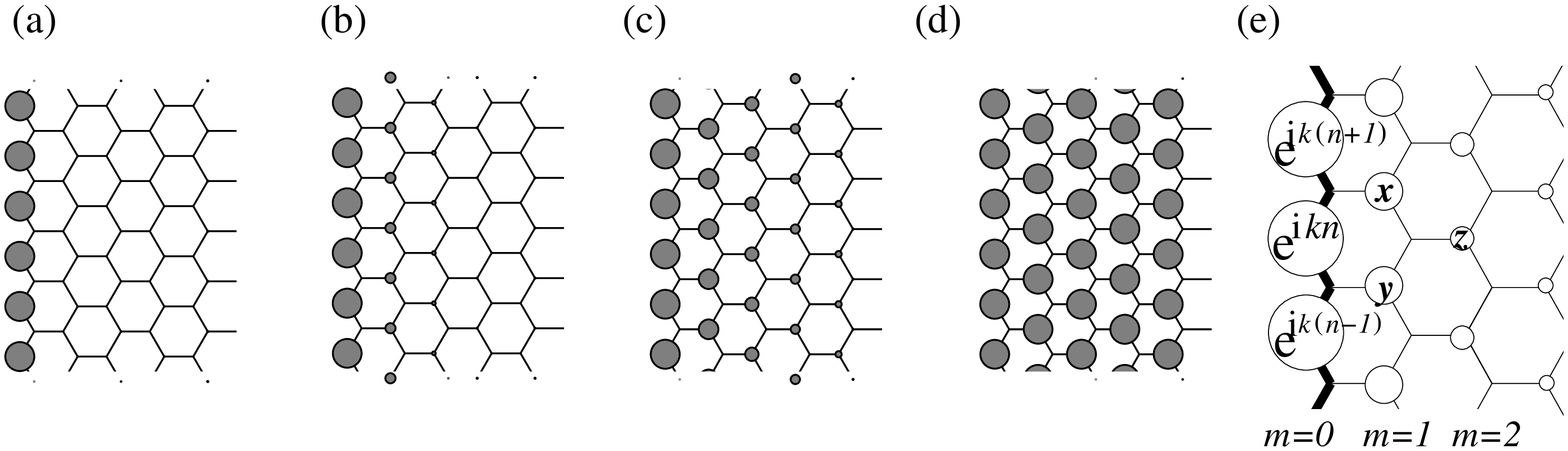}}
\end{center}
\caption{
Charge density plot for analytic solution of
the edge states in a semi-infinite graphite, when (a)$k\! =\! \pi$,
(b)$8\pi /9$, (c)$7\pi /9$ and (d)$2\pi /3$.
(e) An analytic form of the edge state for a semi-infinite graphite sheet
with a zigzag edge, emphasized by bold lines.
Each carbon site is specified by a location index $n$
on the zigzag chain and by a chain order index $m$ from the edge.
The magnitude of the charge density at each site, such as
$x$, $y$ and $z$, is obtained analytically (see text).
The radius of each circle is proportional to the charge density on each
 site,
and the drawing is made for $k=7\pi/9$.
}
\label{fig:analy}
\end{figure}

For zigzag ribbons, however, a remarkable feature arises in the
band structure, as shown in Figs.\ref{fig:zigzag} (a)-(c).
We see that the highest valence band and lowest conduction
band are always degenerate at $k=\pi$.
It is found that the degeneracy of the center bands at $k=\pi$ does
not originate from the intrinsic band structure of graphene sheet.
These two special center bands get flatter with 
increasing ribbon width.
A pair of partial flat bands appears within the region
of $2\pi/3 \le |k|\le \pi$, where the bands sit in the vicinity of
the Fermi level.

The electronic state in the partial flat bands of the zigzag ribbons 
can be understood as the localized state near the zigzag edge 
via examining the charge density
distribution\cite{peculiar,nakada,prb.1999,enoki,fukuyama}. 
Here we show that the puzzle for the emergence of the edge state can
be solved  by considering a semi-infinite graphite sheet
with a zigzag edge.
First to show
the analytic form, we depict the distribution of charge density in the
flat band states for some wave numbers in Fig.\ref{fig:analy}(a)-(d), 
where the amplitude is proportional to the radius.  The
wave function has non-bonding character, {\it i.e.} 
finite amplitudes only on one of the two sublattices which includes the
edge sites.  It is completely localized at the edge site when
$k\!=\!\pi$, and starts to gradually penetrate into the inner sites as
$k$ deviates from $\pi$ reaching the extend state at $k\!=\!2\pi/3$.

Considering the translational symmetry, we can start
constructing the analytic solution for the edge state by
letting the Bloch components of the linear combination
of atomic orbitals (LCAO) wavefunction be
$...$, ${\rm e}^{{\rm i}k(n-1)}$, ${\rm e}^{{\rm i}kn}$, 
${\rm e}^{{\rm i}k(n+1)}$,$...$ on successive edge sites,
where $n$ denotes a site location on the edge.
Then the mathematical condition necessary for the wave function to be
exact for $E=0$ is that the total sum of the components
of the complex wave function over the nearest-neighbor sites should
vanish.
In Fig.\ref{fig:analy}(e), the above condition is 
${\rm e}^{{\rm i}k(n+1)} + {\rm e}^{{\rm i}kn} + x = 0$,
${\rm e}^{{\rm i}kn} + {\rm e}^{{\rm i}k(n-1)} + y = 0$ and
$x+y+z = 0$.
Therefore, the wave function components $x$, $y$ and $z$ are found
to be 
$D_k{\rm e}^{{\rm i}k(n+1/2)}$,
$D_k{\rm e}^{{\rm i}k(n-1/2)}$,
$D_k^2{\rm e}^{{\rm i}kn}$, respectively.
Here $D_k = -2\cos(k/2)$.
We can thus see that the charge density is proportional to
$D_k^{2(m-1)}$ at each non-nodal site of
the $m$-th zigzag chain from the edge.
Then the convergence condition of $|D_k|\le 1$ is required,
for otherwise the wave function would diverge in a semi-infinite
graphite sheet. This convergence condition defines the region
$2\pi/3\le |k|\le \pi$ where the flat band appears.

\subsection{massless Dirac equation}
We briefly discuss here the relation between massless Dirac spectrum
of graphene and low-energy electronic states of nanoribbons.
The electronic states near the two non-equivalent Dirac points
($\bm{K_\pm}$) can be described by
$4\times 4$ Dirac equation, i.e.
\begin{equation}
H_{\bm{k\cdot p}} \bm{F}(\bm{r})
= \epsilon \bm{F}(\bm{r})
\label{kp.eq}
\end{equation}
with
\begin{equation}
H_{\bm{k\cdot p}} 
=
\left(
\begin{array}{cccc}
0 & \gamma (\hat{k}_x-i\hat{k}_y) & 0 & 0  \\
\gamma (\hat{k}_x+i\hat{k}_y) & 0 & 0 & 0  \\
0 & 0 & 0 &  \gamma (\hat{k}_x+i\hat{k}_y) \\
0 & 0 & \gamma (\hat{k}_x-i\hat{k}_y) & 0  \\
\end{array}
\right)
\end{equation}
and 
\begin{equation}
\bm{F}(\bm{r})=\left(
\begin{array}{c}
F_A^{\bm K_+}(\bm{r}) \\
F_B^{\bm K_+}(\bm{r}) \\
F_A^{\bm K_-}(\bm{r}) \\
F_B^{\bm K_-}(\bm{r}) \\
\end{array}
\right).
\end{equation}
Here, $\hat{k}_x$($\hat{k}_y$) is wavevector operator, and can
be replaced as $\bm{\hat{k}}\rightarrow -i\hat{\nabla}$ in the absence of magnetic field. 
$\gamma$ is a band parameter which satisfies $\gamma=\sqrt{3}ta/2$.
$F_A^{\bm K_\pm}(\bm{r})$ and $F_B^{\bm K_\pm}(\bm{r})$ are 
the envelope functions near $\bm{K_\pm}$ points for A and B sublattice which
slowly vary in the length scale of the lattice constant. 
We can rewrite the above effective mass Hamiltonian by 
using the Pauli matrices $\tau^{x,y,z}$ for valley space
($\bm{K}_{\pm}$) as 
\begin{equation}
H_{\bm{k\cdot p}}= \gamma
\left[
\hat{k}_x(\sigma^x\otimes \tau^0 )
+\hat{k}_y(\sigma^y\otimes\tau^z)
\right].
\end{equation}
Here, 
$\tau^0 $ is the $2 \times 2 $ identity matrix.
We can easily obtain the linear energy spectrum for graphene as
\begin{equation}
\epsilon=s\gamma |k|   \qquad  with  \qquad s=\pm 1,
\end{equation}
and the corresponding wavefunctions with the definition of
$\Phi_{\bm{K_\pm}}=\left[F_{\bm{K_\pm}A},F_{\bm{K_\pm}B}\right]$
are
\begin{equation}
\Phi_{\bm{K_\pm}} = 
\frac{1}{\sqrt{2}}
\left(
\begin{array}{c}
s \\
{\rm e}^{\pm i\phi_{\bm k}}
\end{array}
\right)
{\rm e}^{i\bm{k\cdot r}}
\end{equation}
Here
\begin{equation}
{\rm e}^{\pm i\phi_{\bm k}} = \frac{k_x\pm ik_y}{|k_x+ik_y|}.
\end{equation}

\subsubsection{Zigzag nanoribbons}
The low-energy electronic states for zigzag
nanoribbons also can be described starting from the Dirac
equation.\cite{Phd,luis}  
Since the outermost sites along $1^{st}$ ($N^{th}$) zigzag chain
are B(A)-sublattice, an imbalance between two sublattices occurs
at the zigzag edges leading to 
the boundary conditions
\begin{equation}
\phi_{\bm{K_\pm}A}(\bm{r}_{[0]})=0, \quad \phi_{\bm{K_\pm}B}(\bm{r}_{[N+1]})=0,
\end{equation}
where $\bm{r}_{[i]}$ stands for
the coordinate at $i^{th}$ zigzag chain.
The energy eigenvalue and wavenumber is given by the following
relation,
\begin{equation}
\varepsilon = \pm (\eta -k){\rm e}^{\eta W},
\end{equation}
where $\eta=\sqrt{k^2-\varepsilon}$.
It can be shown that
the valley near $k=3\pi/2a$ in Fig.1(b) originates
from the $\bm{K_+}$-point, the other valley at $k=-3\pi/2a$
from $\bm{K_-}$-point.\cite{Phd,luis}

\subsubsection{Armchair nanoribbons}\label{section.arm.wf}
The boundary condition of armchair nanoribbons projects
$\bm{K_+}$ and $\bm{K_-}$ states into $\Gamma$ point in
the first Brillouin Zone as can be seen in Fig.\,\ref{fig:gsheet}(b). 
Thus, the low-energy states for armchair nanoribbons are 
the superposition of $\bm{K_+}$ and $\bm{K_-}$ states. 
The boundary condition for armchair nanoribbons~\cite{luis} can
be written as
\begin{eqnarray}
&[F_{A}^{+}(x,y) + F_{A}^{-}(x,y)]|_{x=0,W} = 0,\\
&[F_{B}^{+}(x,y) - F_{B}^{-}(x,y)]|_{x=0,W} = 0. 
\end{eqnarray}
If the ribbon width $W$ satisfies the condition of $W =(3/2)(N_w+1)a$
with $N_w = 0,1,2,\dots$,  
the system becomes metallic with the linear spectrum. 
The corresponding energy is given by
\begin{equation}
 \epsilon_{n,k,s} = s \gamma \sqrt{\kappa_{n}^{2}+k^{2}}, 
\label{eigenenergy}
\end{equation}
where $\kappa_{n} = \frac{2\pi n}{3(N_w+1)a}$, 
$n = 0, \pm 1,\pm 2, \dots$ and $s = \pm$.
The $n=0$ mode is the lowest linear subband for metallic armchair
ribbons.
The energy gap ($\Delta_s$) to first parabolic subband of $n=1$ is given
as 
\begin{equation}
\Delta_s = 4\pi \gamma/3(N_w+1)a,
\end{equation}
which is inversely proportional to ribbon width.
It should be noted that small energy gap can be acquired due to the
Peierls distortion for half-filling at low temperatures~\cite{igami,son2}, but
such effect is not relevant for single-channel transport in the doped energy
regime. 

\subsection{Edge boundary condition and intervalley scattering}\label{sec.intervalleyscattering}
Now we discuss the relation between the intervalley scattering and
edge boundary condition. 
According to Ref.\,\cite{ando.nakanishi}, the impurity potential can be
included in the massless Dirac equation by adding the following potential term 
$\hat{U}_{\rm imp}$ described as
\begin{equation}
    \hat{U}_{\rm imp}
  = \left( \begin{array}{cccc}
               u_{A}({\bm r}) & 0 & {u'}_{A}({\bm r}) & 0 \\
               0 & u_{B}({\bm r}) & 0 & -{u'}_{B}({\bm r}) \\
               {u'}_{A}({\bm r})^{*} & 0 & u_{A}({\bm r}) & 0 \\
               0 & -{u'}_{B}({\bm r})^{*} & 0 & u_{B}({\bm r}) \\
           \end{array}
    \right),
\end{equation}
with 
\begin{eqnarray}
 u_{X}({\bm r}) & = \sum_{{\bm R}_{X}} g \left(\bm{r}-{\bm R}_{X}\right)
                     \tilde{u}_X\left({\bm R}_{X}\right)\,,
     \\
    \label{eq:u_inter}
 {u'}_{X}({\bm r}) & = \sum_{{\bm R}_{X}} g \left(\bm{r}-{\bm R}_{X}\right)
                     {\rm e}^{-{\rm i}2 {\bm K}\cdot {\bm R}_{X}}
                     \tilde{u}_X\left({\bm R}_{X}\right)\,,
\end{eqnarray}
where $\tilde{u}_{X}(\bm{R}_{X})$ is the local potential due to impurities
for $X = A$ or $B$.
Here $g(\bm{R})$
with the normalization condition of $\sum_{\bm{R}}g(\bm{R})=1$ 
is a real function which has an appreciable amplitude in
the region where $|\bm{R}|$ is smaller than a few times of the 
lattice constant, and decays rapidly with increasing $|\bm{R}|$. 
For convenience we distinguish the impurity into two types by the range
of the impurity potential: 
one is long-ranged impurities (LRI)
if the range of impurity potential is much larger than
the lattice constant and the other is 
short-ranged impurity (SRI) 
if the range of impurity is smaller than the lattice constant.

If only the LRI are present, we can approximate
$u_{A}({\bm r}) = u_{B}({\bm r}) \equiv u({\bm r})$ and
${u'}_{A}({\bm r}) = {u'}_{B}({\bm r}) \equiv {u'}({\bm r})$.
In the case of carbon nanotubes and zigzag nanoribbons, ${u'}_{X}({\bm r})$,
${u'}_{X}({\bm r})$
vanishes after the summation over $\bm{R}_{X}$ in Eq.~(\ref{eq:u_inter})
since the phase factor ${\rm e}^{-{\rm i}2 {\bm K}\cdot {\bm R}_{X}}$
strongly oscillates in the $x$-direction.
This means that the two valleys are independent
and one can only focus on either ${\bm K_+}$ or ${\bm K_-}$ valley.
{\it Thus LRIs do not induce the intervalley scattering for zigzag
nanoribbons.}

However, this cancellation is not complete in an armchair nanoribbon because
the averaging over the $x$-direction is restricted to the finite width of $W$.
This means that we cannot neglect the contribution 
from scatterers particularly in the vicinity of the edges
to ${u'}_{X}({\bm r})$.
{\it This means that
intervalley scattering does not vanish even in the case
of LRI in the armchair nanoribbons.}

\section{Electronic transport properties}
We numerically discuss the electronic transport properties of
the disordered graphene nanoribbons. 
In general, electron scattering in a quantum wire is described by the
scattering matrix.\cite{beenakker.rmt}
Through the scattering matrix $\bm{S}$, the
amplitudes of the scattered waves $\bm{O}$ are
related to the incident waves $\bm{I}$,
\begin{equation}
\left(
\begin{array}{c}
\bm{O_L} \\
\bm{O_R}
\end{array}
\right)
=
\bm{S}
\left(
\begin{array}{c}
\bm{I_L} \\
\bm{I_R}
\end{array}
\right)
=\left(
\begin{array}{cc}
\bm{r} & \bm{t^\prime} \\
\bm{t} & \bm{r^\prime}
\end{array}
\right)
\left(
\begin{array}{c}
\bm{I_L} \\
\bm{I_R}
\end{array}
\right).
\end{equation}
Here, $\bm{r}$ and $\bm{r^\prime}$ are reflection matrices, 
$\bm{t}$ and $\bm{t^\prime}$ are transmission matrices,
$L$ and $R$ denote the left and right lead lines.
The Landauer-B\"uttiker formula\cite{mclbf} relates the
scattering matrix to the conductance of the sample. 
The electrical conductance is calculated using 
the Landauer-B\"uttiker formula,
\begin{equation}
G(E)
= \frac{e^2}{\pi\hbar}{\rm Tr}(\bm{t}\bm{t}^\dagger)
= \frac{e^2}{\pi\hbar} g(E).
\end{equation}
Here the transmission matrix $\bm{t}(E)$ is calculated by means of
the recursive Green function method.\cite{prl,green2}
For simplicity, throughout this paper, we evaluate electronic
conductance in the unit of quantum conductance ($e^2/\pi\hbar$), 
{\it i.e.} dimensionless conductance $g(E)$. 
We would like to mention that recently the edge disorder effect
on the electronic transport properties of graphene nanoribbons
was studied using similar approach.\cite{li,louis,mucciolo}

\subsection{One-way excess channel system}
\begin{figure}[h]
\includegraphics[width=\linewidth]{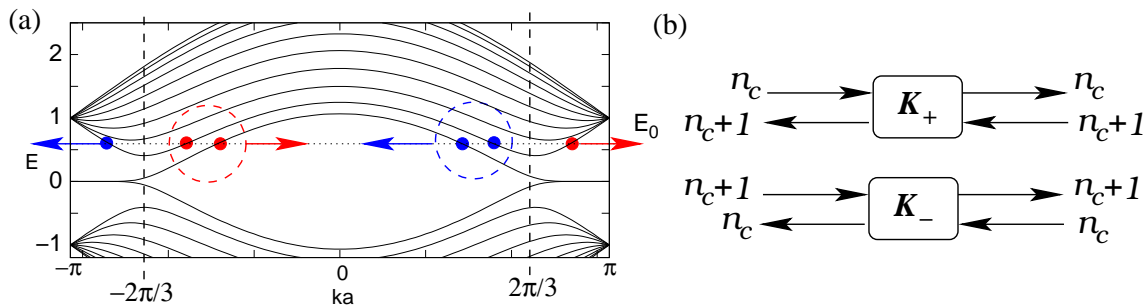}
\caption{
(a) Energy dispersion of zigzag ribbon with $N=10$. The valleys in the 
energy dispersion near $k=2\pi/3a$ ($k=-2\pi/3a$) originate from the
Dirac $\bm{K_+}$($\bm{K_-}$)-point of graphene.
The red-filled (blue-unfilled) circles denote the right (left)-moving
open channel at the energy $E_0$(dashed horizontal line). In the
 left(right) valley, the degeneracy  
between right and left moving channels is missing due to 
one excess right(left)-going mode. The time-reversal symmetry under the
intra-valley scattering is also broken.
(b) Schematic figure of scattering geometry at $\bm{K_+}$ and $\bm{K_-}$
 points in zigzag nanoribbons, where a single excess right-going mode exists
 for $\bm{K_-}$ point. But a single excess left-going mode exists for
 $\bm{K_-}$ point. Here $n_c = 0,1,2,\cdots$.
}
\label{fig:ribbon}
\end{figure}

In this subsection, we consider the conductance of zigzag nanoribbons
in the clean limit, which is simply given by the number of the conducting channel.
As can be seen in Fig. \ref{fig:ribbon}(a), there is always one excess
left-going channel in the right valley ($\bm{K_+}$) within the
energy window of $|E|\le 1$. Analogously, there is one excess
right-going channel in the left valley ($\bm{K_-}$) within the same
energy window. Although the number of right-going and left-going channels
are balanced as a whole system, if we focus on one of two valleys,
there is always one excess channel in one direction, {\it i.e.} a chiral mode.

Now let us consider to inject electrons from left to right-side
through the sample. When the chemical potential is changed from $E=0$, the
quantization rule of the dimensionless conductance ($g_{\bm{K_+}}$) in
the valley of $\bm{K_+}$ is given as
\begin{equation}
g_{\bm{K_+}} = n,
\end{equation}
where $n=0,1,2,\cdots$.
The quantization rule in the $\bm{K_-}$-valley is
\begin{equation}
g_{\bm{K_+}} = n+1.
\end{equation}
Thus, conductance quantization of the zigzag nanoribbon in the clean
limit near $E=0$ has the following odd-number quantization, i.e. 
\begin{equation}
g=g_{\bm{K_+}}+g_{\bm{K_-}}= 2n+1.
\end{equation}

Since we have an excess mode in each valley, the scattering matrix
has some peculiar features which can be seen when we
explicitly write the valley dependence 
in the scattering matrix.
By denoting the contribution of the right valley ($\bm{K_+}$) as $+$,
and of the left valley  ($\bm{K_-}$) as $-$, the scattering matrix
can be rewritten as 
\begin{equation}
\left(
\begin{array}{c}
\bm{O^+_L} \\
\bm{O^-_L} \\
\bm{O^+_R} \\
\bm{O^-_R}
\end{array}
\right)
=\left(
\begin{array}{cc}
\bm{r} & \bm{t^\prime} \\
\bm{t} & \bm{r^\prime}
\end{array}
\right)
\left(
\begin{array}{c}
\bm{I^+_L} \\
\bm{I^-_L} \\
\bm{I^+_R} \\
\bm{I^-_R}
\end{array}
\right).
\end{equation}
Here we should note that the dimension of each column vector is
not identical. Let us denote the number of the right-going channel 
in the valley $\bm{K_+}$ or the left-going channel in the
valley $\bm{K_-}$ as $n_c$. For example,
$n_c = 1$ at $E=E_0$ in Fig.\ref{fig:ribbon}(a).
Fig.\ref{fig:ribbon}(b) shows the schematic figure of scattering
geometry for $\bm{K_+}$ and $\bm{K-}$ points.
Thus the dimension of the column vectors is given as follows:
\begin{equation}
\left\{
\begin{array}{ll}
dim(\bm{I^+_L}) = n_c,     & dim(\bm{I^+_R}) = n_c + 1, \\ 
dim(\bm{I^-_L}) = n_c + 1, & dim(\bm{I^-_R}) = n_c,     \\ 
\end{array}
\right.
\end{equation}
and 
\begin{equation}
\left\{
\begin{array}{ll}
dim(\bm{O^+_L}) = n_c + 1, & dim(\bm{O^+_R}) = n_c,    \\ 
dim(\bm{O^-_L}) = n_c,     & dim(\bm{O^-_R}) = n_c + 1. \\ 
\end{array}
\right.
\end{equation}
Subsequently, the reflection matrices have the following matrix structures, 
\begin{eqnarray}
\bm{r}= \bordermatrix{
      & n_c       & n_c+1         \cr
n_c+1 & \bm{r_{++}} & \bm{r_{+-}} \cr
n_c   & \bm{r_{-+}} & \bm{r_{--}}
},
\end{eqnarray}
\begin{eqnarray}
\bm{r^\prime}= \bordermatrix{
      & n_c+1      & n_c         \cr
n_c & \bm{r^\prime_{++}} & \bm{r^\prime_{+-}} \cr
n_c+1   & \bm{r^\prime_{-+}} & \bm{r^\prime_{--}}
}. 
\end{eqnarray}
The reflection matrices
become non-square when the intervalley scattering is suppressed, {\it i.e.}
the off-diagonal submatrices ($\bm{r_{+-}}$, $\bm{r_{-+}}$ and so on)  are zero.

When the electrons are injected from the
left lead of the sample and the intervalley scattering is suppressed, 
a system with an excess channel is realised in the $\bm{K_-}$-valley.
Thus, for single valley transport,
the $\bm{r_{--}}$ and $\bm{r^\prime_{--}}$ are $n_c\times (n_c+1)$ and $(n_c+1)\times n_c$
matrices, respectively, and $\bm{t_{--}}$ and $\bm{t^\prime_{--}}$ are
$(n_c+1)\times (n_c + 1)$ and $n_c \times n_c$ matrices, respectively.
Noting the dimensions of $\bm{r_{--}}$ and $\bm{r^\prime_{--}}$, we find that
$\bm{r_{--}}^\dagger \bm{r_{--}}$ and $\bm{r^\prime_{--}}
{\bm{{r^\prime}^\dagger_{--}}}$ have a single 
zero eigenvalue. Combining this property with the flux conservation relation
($\bm{S}^\dagger \bm{S} = \bm{S}\bm{S}^\dagger = \bm{1}$), we arrive at
the conclusion that 
$\bm{t_{--}t^\dagger_{--}}$ has an eigenvalue equal to unity, which
indicates the presence of a perfectly conducting channel (PCC)
only in the right-moving channels. 
Note that 
${\bm {t^\prime}_{--}} {\bm {t^\prime}}^\dagger_{--} $
does not
have such an anomalous eigenvalue. 
If the set of eigenvalues for 
${\bm {t^\prime}_{--}} {\bm {t^\prime}}^\dagger_{--}$ is expressed as
$\{T_1, T_2, \cdots, T_{n_c}\}$, that for
${\bm{t_{--}}{\bm t}^\dagger_{\bm --}}$ is
expressed as $\{T_1, T_2, \cdots, T_{n_c}, 1\}$, i.e. a PCC.
Thus, the dimensionless conductance g for the right-moving channels is
given as 
\begin{equation}
g_{\bm{K_-}}=\sum_{i=1}^{n_c+1}T_i = 1 + \sum_{i=1}^{n_c}T_i,
\end{equation}
while that for the left-moving channels is 
\begin{equation}
g^\prime_{\bm{K_-}} = \sum_{i=1}^{n_c}T_i. 
\end{equation}
We see  that $g_{\bm{K_-}}=g^\prime_{\bm{K_-}} +1$.
Since the overall time reversal symmetry (TRS) of the system guarantees
the following relation:
\begin{equation}
\begin{array}{c}
g^\prime_{\bm{K_+}} = g_{\bm{K_-}},\\
g^\prime_{\bm{K_-}} = g_{\bm{K_+}},
\end{array}
\end{equation}
the conductance 
$g=g_{\bm{K_+}}+g_{\bm{K_-}}$ (right-moving) and 
$g^\prime=g^\prime_{\bm{K_+}}+g^\prime_{\bm{K_-}}$ (left-moving) are equivalent.
If  the probability distribution of $\{T_i\}$ is obtained as a function
$L$, we can describe the statistical properties of $g$ as well as
$g^\prime$.
The evolution of the distribution function with increasing $L$ is
described by the DMPK(Dorokhov-Mello-Pereyra-Kumar) equation for 
transmission eigenvalues.\cite{takane4}

In the following, the presence of a perfectly conducting
channel in disordered graphene nanoribbons will be demonstrated 
with the help of numerical calculation. 
Recently Hirose {\it et. al.} pointed out that the
Chalker-Coddington model which possesses non-square 
reflection matrices with unitary symmetry gives rise to a perfectly
conducting channel.\cite{ohtsuki}
However, systems with an excess channel in one direction has been believed difficult to realize. 
Therefore disordered graphene zigzag nanoribbons with LRI  might constitute the first
realistic example. It is possible to extend the discussion 
to generic multiple-excess channel model, where the $m$-PCCs
($m=2,3,\cdots$) appear.\cite{takane4} 
Such systems can be realized by stacking zigzag nanographene
ribbons.\cite{miyamoto}
The electronic transport due to PCC resembles to the electronic
transport due to a chiral mode in quantum 
Hall system.
However, it should be noted that the PCC due to edge states 
in zigzag ribbons occurs even without the magnetic field.\cite{macdonald,ishizaka}

\subsection{Model of impurity potential}
As shown in Fig.1, the impurities are randomly distributed with a density $n_{imp}$ in the
nanoribbons. In our model  we assume that the each impurity potential 
has a Gaussian form of a range $d$
\begin{equation}
V(\bm{r}_i) = \sum_{\bm{r_0}(random)}u
\exp\left(-\frac{|\bm{r}_i-\bm{r}_0|^2}{d^2}\right) 
\end{equation}
where the strength $u$ is uniformly distributed within the range $|u|\le u_{M}$.
Here $u_{M}$ satisfies the normalization condition:
\begin{equation}
u_{M}\sum_{{\bm r}_i}^{(full\ space)}
\exp\left(-{\bm{r}_i^2}{d^2}\right)/(\sqrt{3}/2)=u_0.
\end{equation}
In this work,
we set $n_{\rm imp.}=0.1$, $u_0=1.0$ and 
$d/a=1.5$ for LRI and $d/a=0.1$ for SRI.

\subsection{Perfectly conducting channel: absence of Anderson localization}
\begin{figure}
\includegraphics[width=0.9\linewidth]{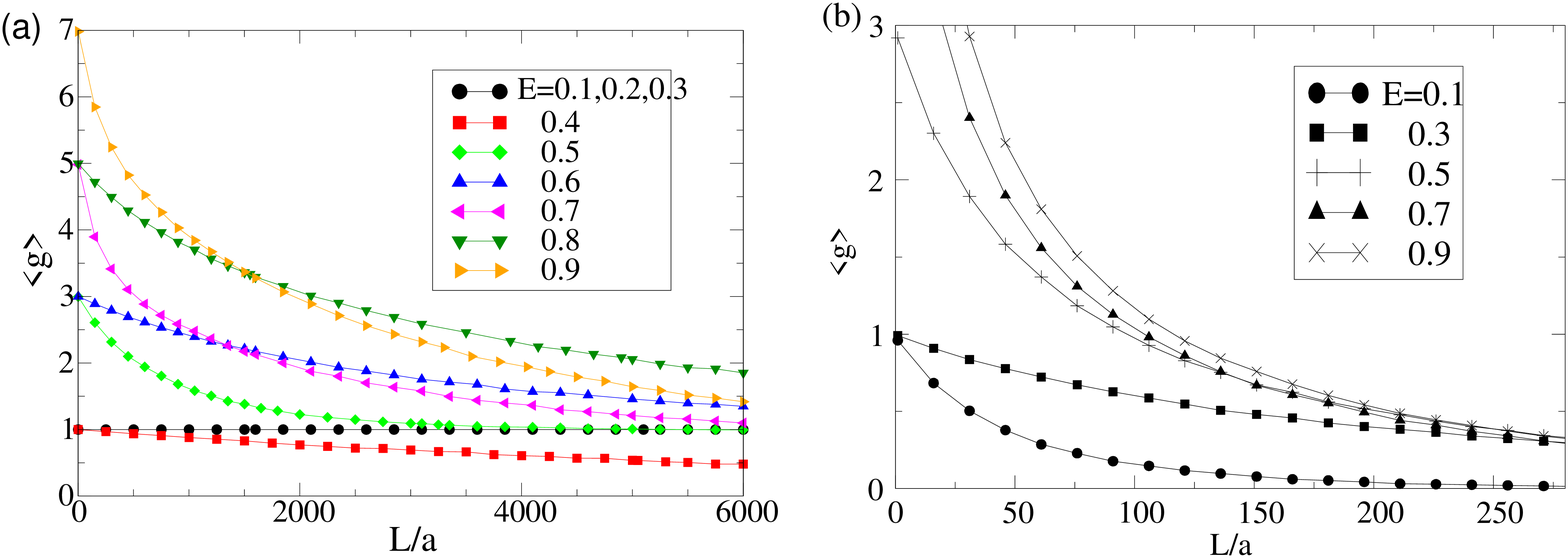}
\caption{$L$-dependence of 
the averaged dimensionless conductance, $ \langle g \rangle $ for zigzag nanoribbon
with $N=10$,  (a) $d/a=1.5$ (no inter-valley scattering), (b)
$d/a=0.1$ (inter-valley scattering). Here $u_0=1.0$, and
$n_{imp.}=0.1$. More than 9000 samples with different impurity configuration
are included in the ensemble average.
}
\label{fig:aveg}
\end{figure}

We focus first on the case of LRI
using a potential with $ d/a=1.5 $ which is already sufficient to avoid inter-valley scattering. 
Fig.\ref{fig:aveg}(a) shows the averaged dimensionless conductance as a
function of $ L $ for different incident energies(Fermi energies), averaging over an
ensemble of 40000 samples with 
different impurity configurations for ribbons of width $N=10$.
The potential strength and impurity density are chosen to be 
$u_0=1.0$ and $n_{imp.} = 0.1$, respectively. As a typical localization effect we observe
that $\langle g \rangle $ gradually decreases with growing length $ L $ (Fig.\ref{fig:aveg}).
However, $ \langle g \rangle $ converges to $\langle g\rangle =1$ for
LRIs (Fig.\ref{fig:aveg}(a)), 
indicating the presence of a single {\it perfectly conducting} 
channel. It can be seen that $\langle g\rangle(L) $ has an exponential behavior as
\begin{equation}
\langle g\rangle -1 \sim \exp(-L/\xi)
\end{equation}
with $\xi$ as the localization length. 

We performed a number of tests to confirm the presence of this perfectly
conducting channel.  
First of all, it exists up to $L=3000a$ for various ribbon widths up to
$N=40$ for the 
potential range ($d/a=1.5$). Moreover the perfectly conducting
channel remains for LRI with
$d/a=2.0,4.0,6.0,8.0$, and $u_0=1.0$,  $n_{imp.}=0.1$ and $ N=10 $. As
the effect is connected with the subtle feature of an excess mode in the
band structure, it is natural that the result
can only be valid for sufficiently weak potentials. For potential strengths comparable to the
energy scale of the band structure, e.g. the energy difference between the transverse modes, 
the result should be qualitatively altered.\cite{vacancy} 
Deviations
from the limit $ \langle g \rangle \to 1 $  
also occur, if the incident energy lies at a value close to the change
between $ g = 2n-1 $ and $ g=2n+1 $  
for the ribbon without disorder. This is for example visible in above
calculations for $E =0.4 $ 
where the limiting value $ \langle g \rangle < 1 $ (Fig.\ref{fig:aveg}(a)). 

Turning to the case of SRI the inter-valley scattering becomes sizable enough
to ensure TRS, such that the perfect transport supported by the
effective chiral mode in a single valley ceases to exist. 
In Fig.\ref{fig:aveg}(b), the nanoribbon length dependence of the
averaged conductance for SRIs is shown.
Since SRI causes the inter-valley scattering for any incident energy, 
the electrons tend to be localized and 
the averaged conductance decays exponentially, $\langle g\rangle\sim \exp(-L/\xi)$, without
developing a perfect conduction channel.

In this subsection, we have completely neglected the effect of
electron-electron interaction, which may acquire the energy gap for
non-doped zigzag nanoribbon at very
low-temperatures accompanying with the edge spin polarization.\cite{peculiar,rpa,son} 
In such situation, small transport gap will appear near $E=0$.
Since the edge states have less Fermi instability for doped regime, the
spin polarized states might be less important for doped system.

\begin{figure*}
\includegraphics[width=0.95\linewidth]{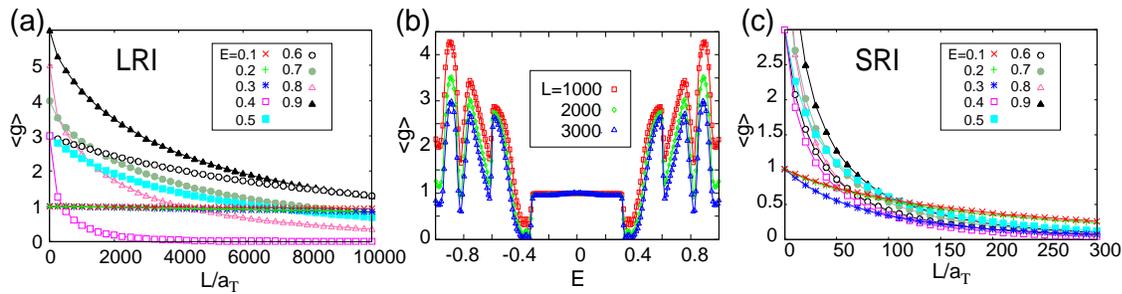}
\caption{(a) Average conductance $\avg{g}$ 
as a function of the ribbon length $L$
in the presence of long-ranged impurities (LRI)
for several different Fermi energies $E$. 
Conductance is almost unaffected by impurities for single-channel
transport ($E=0.1, 0.2 \,\,{\rm and}\,\, 0.3$) while it shows a conventional
exponential decay for multi-channel transport ($E \ge 0.4$).
Here, $N=14$, $n_{\rm imp.}=0.1$ and $d/a=1.5$. 
Ensemble average is taken over $10^4$ samples.
(b) The Fermi energy dependence of $\langle g\rangle$ for LRI. 
(c) The same as (a) for short-ranged impurities (SRI). 
Here, $N=14$, $n_{\rm imp.}=0.1$ and $d/a=0.1$. 
}
\label{fig2}
\end{figure*}

\subsection{Nearly pefect single channel transport in disordered armchair nanoribbons \label{sec3}}
Now we turn to the discussion of the electronic transport properties of
disordered metallic armchair nanoribbons. 
Figure \ref{fig2}(a) shows the averaged conductance $\avg{g}$ as a function of 
the ribbon length $L$ in the presence of LRI for several different Fermi energies $E$.
As we can clearly see, the averaged conductance subjected to LRI in the 
single-channel transport ($E = 0.1, 0.2 \,\,{\rm and}\,\, 0.3$) is
nearly equal to one even in the long wire regime. 
This result is contrary to our expectation that electrons are scattered
even by LRI, 
since wavefunctions at $\bm{K_+}$ and $\bm{K_-}$ points are mixed in
armchair nanoribbons
as we have already seen in Sec.\ref{section.arm.wf}.
For multi-channel transport ($E \ge 0.4$), the conductance shows a conventional decay.
The robustness of single-channel transport can be clearly viewed from the 
Fermi energy dependence of conductance
for several different ribbon lengths $L$ as shown in Fig.\,\ref{fig2}(b).
It should be noted that the energy dependence in the vicinity of $E=0$
is quite different from that in zigzag nanoribbons.
The conductance decays rapidly due to the finite ribbon width
effect in zigzag ribbons~\cite{prl2007} while
the conductance around $E=0$ remains unity 
in armchair ribbons (Fig.\,\ref{fig2}(b)).

Now let us see the effect of short-ranged impurities (SRI).
Figure \ref{fig2}(c) shows the
average conductance $\avg{g}$ as a function of 
the ribbon length $L$ in the presence of SRI
for several different Fermi energies $E$.
In this case, the conductance decays exponentially even for single-channel transport.
This result is similar to that previously obtained in zigzag nanoribbons. 
However, the rate of decay in the low-energy single-channel regime
($E=0.1 \,\,{\rm and} \,\,0.2$) is slower than that for multi-channel
transport regime ($E \ge 0.4$) in this case.  
Similar results are obtained in Ref.~\cite{li}, but in which 
only short-ranged disorder at the edge of ribbon is considered. 

\subsection{$T$-matrix analysis}
The absence of localization in the single-channel region can be 
understood from the Dirac equation including the
impurity potential term $\hat{U}_{\rm imp}$ with armchair edge boundary. 
To consider the amplitude of backward scattering, we introduce
the $T$-matrix defined as
\begin{equation}
  T = \hat{U}_{\rm imp}
      + \hat{U}_{\rm imp}\frac{1}{E - \hat{H}_{0}}\hat{U}_{\rm imp}
      + \cdots .
\end{equation}
We can evaluate the matrix elements of $\hat{U}_{\rm imp}$ for
the eigenstate $| n,k,s \rangle$ with the eigenenergy of
Eq.\,(\ref{eigenenergy}) which can be written as 
\begin{equation}
 | n,k,s \rangle
  = \frac{1}{\sqrt{4WL}}
    \left(
    \begin{array}{l}
      \left( \begin{array}{c}
                 s \\
                 {\rm e}^{-{\rm i}\theta(n,k)}
             \end{array}
      \right) {\rm e}^{{\rm i}\kappa_{n}x}
      \\
      \left( \begin{array}{c}
                 -s \\
                {\rm e}^{-{\rm i}\theta(n,k)}
             \end{array}
      \right) {\rm e}^{-{\rm i}\kappa_{n}x}
    \end{array}
    \right)  {\rm e}^{{\rm i}ky}\,,
\label{wf}
\end{equation}
with the phase factor
\begin{equation}
{\rm e}^{-{\rm i}\theta(n,k)} 
   = \frac{\kappa_{n}-{\rm i}k}{\sqrt{\kappa_{n}^{2}+k^{2}}}\,. 
\end{equation}
Here it should be noted that the phase structure in Eq.\,(\ref{wf}) is 
different between $\bm{K_+}$ and $\bm{K_-}$ states, 
and this internal phase structures are critical for the
scattering matrix elements of armchair nanoribbons
as we discuss in the following.
Using the above expression, we can obtain the 
scattering matrix element
\begin{equation}
     \label{eq:U_matel}
\langle n,k,s | \hat{U}_{\rm imp} | n',k',s' \rangle 
= \left( ss' + {\rm e}^{{\rm i}(\theta(n,k)-\theta(n',k'))} \right) 
V\left(n,k; n^\prime,k^\prime \right)\,,
\label{eq14}
\end{equation}
with
\begin{eqnarray}
V\left(n,k; n', k' \right)
 =  \frac{1}{4WL}\int_{0}^{W}{\rm d}x \int_{0}^{L}{\rm d}y \,
  {\rm e}^{-{\rm i}(k-k')y}\nonumber\\
 \times  
 \left[  u(\bm{r}) \left( {\rm e}^{-{\rm i}(\kappa_{n}-\kappa_{n'})x}
                             + {\rm c.c.} \right) \right.
- \left.\left( u'(\bm{r}){\rm e}^{-{\rm i}(\kappa_{n}+\kappa_{n'})x}
                    + {\rm c.c.} \right) 
    \right] .
\label{eq15}
\end{eqnarray}
It should be emphasized that Eq.\,(\ref{eq14}) has the same form as that
obtained for carbon nanotubes without inter-valley scattering ($u'_X(\bm{r}) = 0$)\,\cite{ando.nakanishi}.
Interestingly, in spite of the fact that armchair nanoribbons inevitably suffer from the inter-valley
scattering due to the armchair edges ($u'_X(\bm{r}) \ne 0$),
we can express the matrix element for the backward scattering as Eq.\,(\ref{eq14})
by including $u'_X(\bm{r})$ into $V(n, k; n', k')$ in Eq.\,(\ref{eq15}).
This is due to the different phase structure between $\bm{K_+}$ and $\bm{K_-}$
in Eq.\,(\ref{wf}).

We focus on the single-channel regime where only the lowest subband
with $n = 0$ crosses the Fermi level.
From Eq.\,(\ref{eq:U_matel}), 
the scattering amplitude from the propagating state 
$| 0,k,s\rangle$ to its backward state
$| 0,-k,s\rangle$ in the single-channel mode becomes
identically zero, {\it i.e.} 
\begin{equation}
  \langle 0,-k,s | \hat{U}_{\rm imp} | 0,k,s \rangle = 0 .
\end{equation}
Thus, since the lowest backward scattering matrix element of $T$-matrix
vanishes, the decay of $\langle g \rangle$ in the single channel energy
regime is extremely slow as a function of the ribbon length as we have
seen in Fig.\,\ref{fig2}.
However, the back-scattering amplitude in the second and much higher order
does not vanish.
Hence the single-channel conduction is not exactly perfect like carbon nanotubes\,\cite{ando.nakanishi},
but {\it nearly} perfect in armchair nanoribbons.

\section{Universality class}
According to random matrix theory, ordinary disordered quantum wires are classified into
the standard universality classes, orthogonal, unitary and symplectic.
The universality classes describe transport properties
which are independent of the microscopic details of disordered wires.
These classes can be specified by time-reversal and spin rotation
symmetry.
The orthogonal class consists of systems having both time-reversal and
spin-rotation symmetries, while the unitary class is characterized by
the absence of time-reversal symmetry.
The systems having time-reversal symmetry without spin-rotation symmetry
belong to the symplectic class.
These universality classes have been believed to
inevitably cause the Anderson localization
although typical behaviors are different from class to class.

In the graphene system, the presence or absence of the intervalley
scattering affect the time reversal symmetry of the system. 
If the inter-valley scattering is absent, 
{\it i.e.} $u'_X(\bm{r})= 0$, 
the Hamiltonian $\hat{H}_{0} + \hat{U}_{\rm imp}$ 
becomes invariant under the transformation of
$\mathcal{S} = -{\rm i}\left(\sigma^{y}\otimes \tau^{0}\right)C$,
where $C$ is the complex-conjugate operator.
This operation corresponds to the special time-reversal operation for pseudospins within
each valley, and supports that the system has the symplectic symmetry.
However, in the presence of inter-valley scattering due to SRI, 
the invariance under $\mathcal{S}$ is broken.
In this case, the time reversal symmetry across two valleys described by
the operator $\mathcal{T}= \left(\sigma^z\otimes\tau^x\right)C$
becomes relevant, which indicates orthogonal universality class.
Thus as noted in Ref.~\cite{suzuura.prl}, 
graphene with LRI belongs to symplectic symmetry, but that with SRI
belongs to orthogonal symmetry. 

However, in the zigzag nanoribbons, the boundary conditions which treat
the two sublattices asymmetrically 
leading to edge states give rise to a single special mode in each valley.
Considering now one of the two valleys separately,
say the one around $k = k_{+}$, we see that the pseudo TRS is violated
in the sense that we find one more left-moving than right-moving mode.
Thus, as long as disorder promotes only intra-valley scattering,
the system has no time-reversal symmetry.
On the other hand, if disorder yields inter-valley scattering,
the pseudo TRS disappears but the ordinary TRS is relevant making
a complete set of pairs of time-reversed modes across the two valleys.
Thus we expect to see qualitative differences in the properties
if the range of the impurity potentials is changed.

The presence of one perfectly conducting channel
has been recently found in disordered metallic carbon nanotubes with LRI.\cite{suzuura}
The PCC in this system originates from the 
skew-symmetry of the reflection matrix, 
$^t\bm{r}= -{\bm r}$,\cite{suzuura}
which is special to the symplectic symmetry
with odd number of channels.
The electronic transport properties such system
have been studied on the basis of the random matrix theory.\cite{takane1, takane2}
On the other hand, zigzag ribbons without inter-valley scattering are
not in the symplectic class, since they break TRS in a special way. 
The decisive feature for a perfectly conducting channel is
the presence of one excess mode in each valley as discussed in the
previous section.

\begin{figure}
\includegraphics[width=0.6\linewidth]{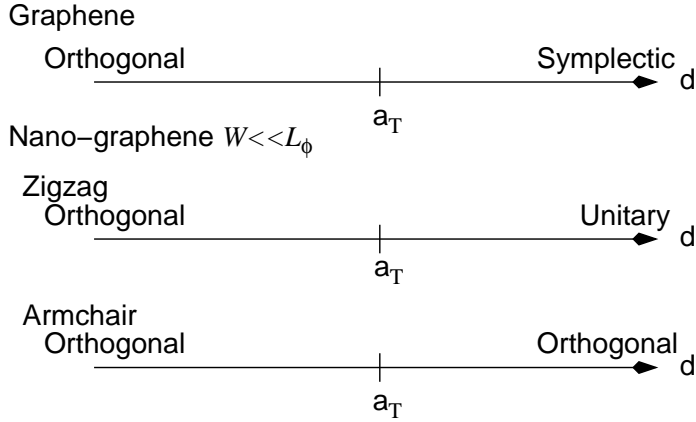}
\caption{Summary concerning the universality crossover. With
 increasing the range of the impurity potential, graphene is known to be
the orthogonal for SRIs and the symplectic class for LRIs. However,
zigzag nanoribbons are unitary class for SRIs. Armchair ribbons are
classified into orthogonal class for all the range of the impurity. $L_\phi$ is the phase
 coherence length. $W$ is the width of graphene ribbons.
}
\label{fig:crossover}
\end{figure}
In view of this classification we find that the universality class of
the disordered zigzag nanoribbon 
with long-ranged impurity potential (no inter-valley scattering) is the {\it unitary} class 
(no TRS). On the other hand, for short-range impurity potentials 
with inter-valley scattering the disordered ribbon belongs to the {\it orthogonal} class 
(with overall TRS). 
Consequently we can observe a crossover between two universality classes when we change
the impurity range continuously.

However, in the disordered armchair nanoribbons,
the special time-reversal symmetry within each valley is broken
even in the case of LRI.
This is because $u'_X(\bm{r}) \ne 0$ as we have seen 
in Section \ref{sec.intervalleyscattering}.
Thus, irrespective of the range of impurities,
the armchair nanoribbons are classified into orthogonal universality class. 
\begin{it}
Since the disordered zigzag nanoribbons are classified
into unitary class for LRI but orthogonal class for SRI~\cite{prl2007}, 
it should be noted that the universality crossover in nanographene
system can occur not only 
due to the range of impurities but also due to the edge boundary
conditions.
\end{it}

\begin{figure}
\includegraphics[width=0.9\linewidth]{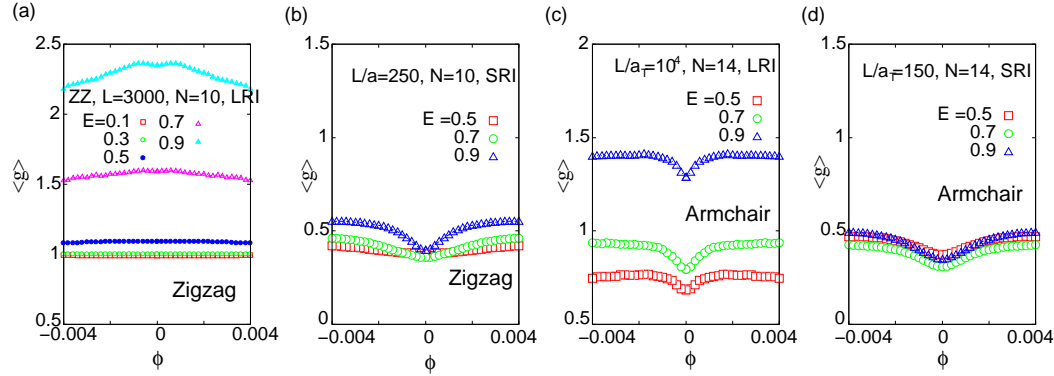}
\caption{The magnetic field dependence of the averaged conductance
for disordered zigzag nanoribbons with (a) LRIs and (b) SRIs.
Similarly, the case for armchair nanoribbons with (c) LRIs and (d) SRIs.
$\phi$ is the magnetic flux through a hexagon ring which is measured in 
units of $ch/e$.}
\label{fig:magneto}
\end{figure}

The application of a magnetic field enforces the above arguments. 
In Fig.\ref{fig:magneto}, the magnetic field dependence of the averaged
conductance for disordered zigzag nanorribons with (a) LRIs and (b)
SRIs is presented.
Similarly, the case for armchair nanoribbons with (c) LRIs and (d) SRIs.
Here we have included the magnetic filed perpendicular to the 
graphite plane which are incorporated via the Peierls phase: 
$\gamma_{i,j}\rightarrow\gamma_{i,j}\exp\left[
i2\pi\frac{e}{ch}\int_i^jd\bm{l\cdot A}\right]$, 
where $\bm{A}$ is the vector potential. 
Since the time-reversal symmetry within the valleys for zigzag ribbons
is already broken, the averaged conductance $\langle g\rangle$
for LRIs (absence of intervalley scattering) is quite
insensitive to the application of magnetic field as can be seen in 
Fig.\ref{fig:magneto}(a). This is consistent with the behavior 
in the unitary class.
For higher energies, weak magnetic field
dependence appears due to the intervalley scattering.
For all the other cases weak magnetic filed improve the conductance,
{\it i.e.} weak localization behavior which is typical for the orthogonal class.

\section{Summary \label{sec4}}
In this paper, we have presented a brief overview on the electronic and transport properties
of graphene nanoribbons focusing on the effect of edge shapes and 
impurity scattering. 
Concerning transport properties for disordered
systems the most important consequence is the presence of a perfectly
conducting channel in zigzag nanoribbons, {\it i.e.} the absence of Anderson localization which is
believed to inevitably occur in the one-dimensional electron system.
The origin of this effect lies in the single-valley transport which is
dominated by a chiral mode. On the other hand, large momentum transfer
through impurities with short-range potentials involves both valleys,
destroying this effect and leading to usual Anderson localization.  The
obvious relation of the chiral mode with time reversal symmetry leads
to the classification into the unitary and orthogonal class 
depending on the range of impurity potential. 
On the other hand, in spite of the lack of well-separated two 
valley structures, the single-channel transport subjected to
long-ranged impurities shows nearly perfect transmission, 
where the backward scattering matrix elements in the lowest order vanish 
as a manifestation of internal phase structures of the wavefunction.
These results are in contrast with the mechanism of perfectly conducting
channel in disordered zigzag nanoribbons and metallic nanotubes
where the well separation between two non-equivalent Dirac points is
essential to suppress the inter-valley scattering. 

\section*{ACKNOWLEDGEMENT}
This work was financially supported by a Grand-in-Aid for Scientific
Research from the MEXT and the JSPS (Nos. 19710082, 19310094, 20001006,
and 21540389).

\end{document}